\documentclass[12pt,a4paper,english,nofootinbib]{revtex4-2}
\usepackage{tcolorbox}
\usepackage{lmodern}
\usepackage{lmodern}

\usepackage[T1]{fontenc}
\usepackage[latin9]{inputenc}
\usepackage{babel}
\usepackage{calc}
\usepackage{amsmath}
\usepackage{amssymb}
\usepackage{graphicx}
\usepackage{esint}
\usepackage[unicode=true,pdfusetitle,
 bookmarks=true,bookmarksnumbered=false,bookmarksopen=false,
 breaklinks=false,pdfborder={0 0 1},backref=false,colorlinks=false]
 {hyperref}
\usepackage{xcolor}
\makeatletter

\@ifundefined{textcolor}{}
{%
 \definecolor{BLACK}{gray}{0}
 \definecolor{WHITE}{gray}{1}
 \definecolor{RED}{rgb}{1,0,0}
 \definecolor{GREEN}{rgb}{0,1,0}
 \definecolor{BLUE}{rgb}{0,0,1}
 \definecolor{CYAN}{cmyk}{1,0,0,0}
 \definecolor{MAGENTA}{cmyk}{0,1,0,0}
 \definecolor{YELLOW}{cmyk}{0,0,1,0}
}

\usepackage{babel}

\usepackage{graphicx}
\def\b{\begin{equation}}
\def\e{\end{equation}}

\@ifundefined{textcolor}{}{%
 \definecolor{BLACK}{gray}{0}
 \definecolor{WHITE}{gray}{1}
 \definecolor{RED}{rgb}{1,0,0}
 \definecolor{GREEN}{rgb}{0,1,0}
 \definecolor{BLUE}{rgb}{0,0,1}
 \definecolor{CYAN}{cmyk}{1,0,0,0}
 \definecolor{MAGENTA}{cmyk}{0,1,0,0}
 \definecolor{YELLOW}{cmyk}{0,0,1,0}
 }

\usepackage{latexsym}\usepackage{bm}
\makeatother

\begin{document}
\title{Mass formulas for individual black holes in merging binaries }
\author{{\normalsize{}{}{}{}{}{}{}{}{}{}{}{}{}{}{}{}{}Zeynep Tugce Ozkarsligil
}}
\email{ozkarsligil.zeynep@metu.edu.tr}

\author{{\normalsize{}{}{}{}{}{}{}{}{}{}{}{}{}{}{}{}{}Bayram
Tekin}}
\email{btekin@metu.edu.tr}

\affiliation{Department of Physics, Middle East Technical University, 06800, Ankara,
Turkey}
\date{{\normalsize{}{}{}{}{}{}{}{}{}{}{{}{}{}{}\today}}}

\date{{\normalsize{}{}{}{}{}{}{}{}{}{}{{}{}{}{}\today}}}

\begin{abstract}

\noindent We give formulas for individual black hole masses in a merger, by using Newtonian physics, in terms of the three measured quantities in the detector: the initial wave frequency $f_1$, the maximum detected frequency (chirp frequency)  $f_2$, and the time elapse $\tau$ between these two frequencies. Newtonian gravity provides an excellent pedagogical tool to understand the basic features of gravitational wave observations, but it must be augmented with the assumption of gravitational radiation from General Relativity for accelerating masses  as there is no gravitational wave in Newtonian gravity.  The simplest approach would be to consider a binary system of two non-spinning masses (two black holes) circling their common center of mass. All the computations can be done within Newtonian physics, but the General Relativistic formula for the power carried by gravitational waves is required in this scheme. It turns out there is a subtle point: for the consistency of this simple, yet pedagogical computation, taking the lowest order power formula from  General Relativity leads to complex individual masses. Here we remedy this problem and {suggest} a way to write down an average power formula coming from perturbative General Relativity.
\end{abstract}

\maketitle

\section{Introduction}

Black hole collision and the physics of the ensuing gravitational wave emission, in its full nonlinear form, is highly nontrivial; and allows only approximate analytical understanding in the inspiral and ringdown phases. The crucial merger phase, when the amplitude of the gravitational wave is the largest, is left to the numerical relativity computations. There is of course no problem with this state of affairs,  but it would be quite useful to have some understanding of the merger and the gravitational radiation phenomena using basic introductory physics. Fortunately, two such beautiful expositions \cite{Mathur,LIGO_team}, appeared where non-relativistic physics is used to describe the merger and estimate the source parameters using the measured data. Of course, as Newtonian theory does not allow gravitational waves by fiat, this input must be taken from General Relativity (GR).  So the only input from GR is the  {energy loss per unit time } (power) formula due to gravitational radiation. The papers \cite{Mathur,LIGO_team} give a good understanding of the measured gravitational wave data in terms of interacting and merging binaries.  In the detectors, strain (relative change in the distances of the detector mirrors), frequency change in the transient wave, and the total time in the detector frame are measured. All the information about the source, such as the luminosity distance of the source, the interacting masses, the energy lost to gravitational waves, etc., can be inferred using these data and the underlying theory. Of course, if the underlying theory is Newtonian gravity, one should only expect an order-of-magnitude matching, not an accurate one.  

In almost all the popular talks given by the second author regarding gravitational wave observations, a persistent question arises that is: how can one determine the individual masses of the merging black holes? Of course, the correct answer can be obtained from full numerical GR as stated above, hence there is no simple analytical formula for the individual masses in terms of the observables or measured quantities in the detector. But, in the context of Newtonian physics, we would like to remedy this here and give the masses of the merging black holes in terms of the measured quantities in the detector. In no way this formula for the masses can substitute actual, numerical calculations; it is just a heuristic discussion that aims to provide students and teachers with a cursory understanding of the merging black holes.   We hope that this will be a nice amendment to the \cite{Mathur,LIGO_team}. See the main formulas as equation ({\ref{main_mass}). The crucial point will be this: the lowest-order gravitational power formula leads to complex individual masses, not real ones, so one has to consider an effective power law that takes into account higher-order corrections in GR.

\section{Individual Masses of Black Holes from Newtonian Physics}

Let us consider two {\it non-spinning}  black holes with masses $m_1$ and $m_2$ moving in circular orbits about their common center of mass.  The total mass is $m_T= m_1+ m_2$ while the reduced mass is $\mu = \frac{m_1 m_2}{m_T}$. The Newtonian equation of motion is
\begin{equation}
\mu \ddot{\bf{r}}  = - \frac{ G \mu m_T}{r^3}{\bf{r}},
\end{equation}
with the relative position given as ${\bf{r}} = {\bf{r}}_2- {\bf{r}}_1$.  Then the orbital angular velocity reads
\begin{equation}
\omega_o^2 =  \frac{ G m_T}{r^3}.
\label{omega_eqn}
\end{equation}
In this circular orbit, the total energy of the system is
\begin{equation}
E_T =\frac{1}{2} \mu v^2 - \frac{G \mu m_T}{r} = - \frac{G \mu m_T}{2 r} = 
- \frac{\mu}{2}\left ( G m_T \omega_o \right )^{2/3},
\label{uc}
\end{equation}
where in the second and third equalities in (\ref{uc}),  we used $v= \omega_o r$ with $\omega_o$ given in (\ref{omega_eqn}). The power loss due to gravitational radiation can be defined as negative of the rate of change in the total energy as 
\begin{equation}
P_\text{loss} = - \frac{ d E_T}{ d t} =  \frac{\mu}{3}\left ( G m_T \right )^{2/3} \omega_o^{-1/3} \frac {d \omega_o}{d t}.
\label{pow}
\end{equation}
Needless to say, even though one can define and calculate the power loss formula as above, gravitational radiation emission does not happen in Newtonian theory. {So, there really is no power loss, and the circular orbit is stable;} but having learned from GR that accelerating mass and energy radiate, we must introduce this concept to Newtonian theory by hand to proceed.}{Then the next question is the following: given two orbiting masses as above, what could be the emitted power formula?  One can take guidance from electrodynamics for which one has electromagnetic field radiation from an oscillating electric dipole. It turns out that in gravity, due to the absence of negative masses, there are no gravitational dipoles, and the leading quantity that is responsible for gravitational radiation is the time-dependent quadrupole moment of the accelerating masses. Therefore the quadrupole moment should appear in the power formula. With this knowledge,} as argued in \cite{Mathur}, a simple argument and dimensional analysis lead us to the power loss formula up to a constant $\alpha$ via gravitational waves as
\begin{equation}
P_\text{GW}= \alpha \frac{ G I^2 \omega_o^6}{ c^5}.
\label{power_eqn}
\end{equation}
Here $I = \mu r^2 $ is the moment of inertia which is time-dependent since $r$ is time-dependent; and $\alpha = 32/5$
at the {\it lowest} (linear) order \cite{Weinberg}. We shall keep this {constant} as $\alpha$, since, as we will show below, the linearized value is not sufficient to describe the LIGO data. Namely, it leads to an inconsistency, {\it i.e.} it will yield complex individual masses of black holes.  We can recast (\ref{power_eqn}) in such a way that the time-dependence is only on {the orbital frequency} $\omega_o$:
\begin{equation}
P_\text{GW}= \alpha \frac{ G^{7/3} \mu^2 m_T^{4/3} \omega_o^{10/3}}{ c^5}.
\label{power_eqn2}
\end{equation}
{Using (\ref{power_eqn}) and  (\ref{power_eqn2}) in  $P_\text{loss} = P_\text{GW}$,}  one arrives at
\begin{equation}
\mu m_T^{2/3} = \frac{ c^5}{3 \alpha G^{5/3} } \omega_o^{-11/3} \frac {d \omega_o}{d t},
\end{equation}
which is usually written in terms of the so-called chirp mass \cite{LIGO_disco}
{\begin{equation}
 {\mathcal{M}}_c := \frac{( m_1 m_2)^{3/5}}{ (m_1+ m_2)^{1/5}} =\mu^{3/5}m_T^{2/5}, 
 \label{ivmehakem}
 \end{equation}}
as
\begin{equation}
{\mathcal{M}}_c = \frac{c^3}{G} \left (\frac{1}{ 3 \alpha} \omega_o^{-11/3} \frac {d \omega_o}{d t} \right)^{3/5}.
\label{chirp_eqn}
\end{equation}
This tells us the time-dependence of the chirp mass during the merger, which does not change much {\cite{LIGO_team}}. Hence we can assume it to be constant and integrate (\ref{chirp_eqn}) to find the finite change in the orbital frequency over a time interval of $\tau$. But before we do that, let us relate the orbital angular frequency $\omega_o$ of the source to the frequency $f$ measured in the detector or the frequency of the gravitational wave at the detector. The relation is as follows: $2 \pi f= 2 \omega_o$. {Note that factor 2 on the right-hand side can be understood by explicitly writing the emitted power in terms of the time dependence of the quadrupole moment. This can also be understood from the fact that the gravitational field has spin-2, namely as the binary system rotates at an angle of $2 \pi$, the emitted gravitational field rotates by an angle $4 \pi$.  }

{There is another assumption we have made: the gravitational source is not far away from our detectors. If this is not the case, we should also consider the redshift of the gravitational wave due to the expansion of the universe. This red shift of the gravitational wave is analogous to the usual cosmological redshift of light coming from large cosmological distances.  The distances are conventionally measured by a dimensionless quantity, the so-called $z$-factor defined as $z:= \frac{f_{\text emitted}}{f_{\text observed}} -1$. } High $z$ values refer to large distances. 
Our assumption of a close source is valid because, in the first merger paper \cite{LIGO_disco}, the merging black holes were not far away, indeed   $z-$factor is just 0.09 \cite{LIGO_disco}. But if one wants to incorporate the redshift of the gravitational wave, one should set $f= \frac{\omega_o}{ \pi (1+ z)}$.  So then, from (\ref{chirp_eqn}), one has the chirp mass in terms of the measured quantities as 
\begin{equation}
{\mathcal{M}}_c=\frac{c^3 }{G \left(8 \pi ^{8/3} \alpha  \tau \right)^{3/5}}\left(\frac{1}{f_1^{8/3}}-\frac{1}{f_2^{8/3}}\right)^{3/5},
\label{chirp_denk}
\end{equation}
where $f_1$ is the initial frequency detected and $f_2 > f_1$ is the final one after a total recording time $\tau$.\footnote{In \cite{Mathur}, $f_2$ is taken to be infinite for the estimation of the chirp mass. Here we kept it just to get a better numerical estimate.}  The individual masses of black holes  from the definition of the chirp mass {(\ref{ivmehakem})}  and the total mass follows as
\begin{equation}
m_{1,2} = \frac{m_T}{2} \Bigg ( 1 \pm  \sqrt{ 1 - 4  \left(\frac{{\mathcal{M}}_c}{m_T}\right)^{5/3}} \Bigg).
\label{mass_exact}
\end{equation}
It is clear that for this formula to be valid, one must have
$m_T \ge  2^{6/5} {\mathcal{M}}_c \approx 2.2974 {\mathcal{M}}_c $.  For the first LIGO observation \cite{LIGO_disco}, these masses were given as $m_T = 70 M_\odot$ and ${\mathcal{M}}_c = 30 M_\odot$; and since for these LIGO values $m_T =2.3333 {\mathcal{M}}_c $, the formula (\ref{mass_exact}) is valid as expected.

Since we want to write $m_{1,2}$ in terms of the {\it measured } quantities, we need to figure out what the total mass $m_T$  is in terms of these quantities. For this purpose, we can assume that \cite{Mathur}, the merger stops once a Schwarzschild black hole with total mass $m_T$ is formed, and at that moment the largest frequency wave with frequency $f_2$ is emitted. So here we assume that gravitational radiation carries only a small fraction of the initial total mass, and we also do not consider the ring-down phase,  {that is the damped oscillations of the single black hole formed after the merger which also leads to a generation of gravitational waves. These assumptions, save the nonspinning assumption are rather reasonable assumptions. We cannot incorporate the spin within the simple Newtonian framework.   The Schwarzschild black hole with mass $m_T$ has the radius $r$ (that is the location of the event horizon)  given as $r=  \frac{ 2 m_T G}{ c^2}$; therefore }
\begin{equation}
m_T = \frac{ c^2 r}{ 2 G} = \frac{c^3}{2\sqrt{2}\pi f_2 G}, 
\label{geyik}
\end{equation}
where in the second equality we used (\ref{omega_eqn}). Plugging the last equation and (\ref{chirp_denk}) in  (\ref{mass_exact}), one arrives at the desired formula for the masses of individual black holes { in terms of what is observed in the detector: the observation time $\tau$, the initial frequency $f_1$ and the final frequency $f_2$.  }
\begin{tcolorbox}
\begin{equation}
m_{1,2}= \frac{c^3}{4 \sqrt{2} \pi  f_2 G}\left(1 \pm \sqrt{1-\frac{2 \sqrt{2} f_2^{5/3}}{ \pi \alpha \tau}\left(\frac{1}{f_1^{8/3}}-\frac{1}{f_2^{8/3}}\right)}\,\, \right).\,\,\,\,\,
\label{main_mass}
\end{equation}
\end{tcolorbox}
\noindent The result will be in kilograms. This formula is valid under the following condition that makes the square root real
\begin{equation}
\alpha \tau \ge \frac{2 \sqrt{2} f_2^{5/3}}{ \pi }\left(\frac{1}{f_1^{8/3}}-\frac{1}{f_2^{8/3}}\right).
\label{bilal1}
\end{equation}
Let us now apply this to the first LIGO observation \cite{LIGO_disco}, one can read the following frequencies $f_1 = 42$ Hz, $f_2= 300$ Hz from the published frequency versus time graph. Plugging these in the right-hand side of (\ref{bilal1}),  one obtains the condition $\alpha \tau \ge 0.5649$. From the same data one can estimate the time it takes the frequency to increase from $f_1$ to $f_2$ to be  $\tau = 0.08$ s, then {the condition (\ref{bilal1})} requires $\alpha \ge 7.0613$. That means, if one naively takes $\alpha$ to have the lowest order value coming from GR, that is  $\alpha = 32/5=6.4$, then $m_{1,2}$ {as computed from (\ref{main_mass})} will be complex numbers. In fact, in \cite{Mathur},  $\alpha = 32/5$, together with the other estimates for $\tau$, $f_1$, and $f_2$ quoted above were used. Then one {obtains }
$m_{1,2} = (38.07 \pm 12.27 i) M_\odot$, which are nonphysical. 

The all-important conclusion one should derive from the above result is that to be able to explain the black hole merger data in a somewhat semi-Newtonian analysis, one cannot just take the lowest order term borrowed from GR  in the radiated-power formula (\ref{power_eqn}). One has to go beyond the first order in perturbation theory to obtain physically viable masses. But, then the ensuing discussion becomes rather complicated since $\alpha$ in (\ref{power_eqn}), instead of being the constant $32/5$, becomes a non-linear function of the involved masses and the source frequency. The resulting radiated-power formula can only be computed in perturbation theory. Such a computation is difficult and is beyond the scope of this paper, but if the reader accepts the perturbative formula known in the literature, and is given below, then the ensuing discussion can be easily followed. 

Let us first note that perturbative corrections to the usual Newtonian potential in GR are often called {\it Post-Newtonian (PN)} corrections; and the  PN formalism is a powerful tool to compute the backreaction effects of the emitted gravitational radiation on the source especially in the near-zone approximation, that is when one is interested in the physics of gravitational waves close to the gravitating object. Below we will use the power law at the so-called 3.5 PN  order. The designation 3.5 PN here refers to the approximation that one considers all the corrections to the relevant physical quantities {\it up to and including} order $\left (\frac{v^2}{c^2}\right)^{3.5}$ corrections where $v$ is the speed of the source that depends on other physical parameters such as the masses and the relevant distances in the problem and $c$ is the speed of light. 

To be able to express the 3.5 PN order correction to the power law in terms of physical variables, and quote the radiated power formula in a merging binary up to 3.5PN order 
\cite{Blanchet,Dam_baba}, as given by equation (5.257) in \cite{Maggiore}, let us define two dimensionless quantities 
\begin{eqnarray}
 x:= \left( \frac{G m_T \omega_o}{c^3}\right)^{2/3}, \hskip 1 cm \nu := \frac{m_1 m_2}{ m_T^2}.
 \label{bluh}
\end{eqnarray}
The order of  $x$  can be estimated to be $x = {\mathcal{O}}(\frac{v^2}{c^2})$, so it a priori takes values as $0 \le x <1$, while $\nu$ takes values in the interval $0 \le \nu \le 1/4$,  and the upper bound is satisfied when the merging masses are equal.
Then up to and including {${\mathcal{O}}(\frac{v^7}{c^7})$, the power formula reads
\begin{eqnarray}
P_{GW}= \frac{32 c^5}{5 G}\nu^2 x^5 g(x,\nu) =: \alpha_{eff}(x,\nu)\frac{c^5}{G}\nu^2 x^5, 
\label{salakmeral}
\end{eqnarray}
where the dimensionless function is
\begin{eqnarray} 
g(x,\nu) := &&1 - \left(\frac{35 \nu }{12}+\frac{1247}{336}\right) x + 4 \pi x^{3/2}+\left(\frac{65 \nu ^2}{18}+\frac{9271 \nu }{504}-\frac{44711}{9072}\right) x^2 \nonumber \\
&& - \left(\frac{583 \nu }{24}+\frac{8191}{672}\right)\pi x^{5/2} +\left(\frac{193385 \nu ^2}{3024}+\frac{214745 \nu }{1728}-\frac{16285}{504}\right) \pi x^{7/2}  \\
&&+x^3 \Bigg (-\frac{1712 \gamma_E }{105}-\frac{775 \nu ^3}{324}-\frac{94403 \nu ^2}{3024}+\left(\frac{41 \pi ^2}{48}-\frac{134543}{7776}\right) \nu \nonumber \\
&&-\frac{856}{105} \log (16 x)+\frac{16 \pi ^2}{3}+\frac{6643739519}{69854400}\Bigg).\nonumber
\label{cenabettin}
\end{eqnarray}
Here $\gamma_E = 0.577..$ is the {Euler-Mascheroni} constant. At the lowest order, one takes $g(x,\nu)=1$, but as we argued above, this does not yield consistent results for individual masses. One way to take into account this non-trivial dependence of $g(x,\nu)$ on $x$ is to consider its average value from initial $x_1$ to the final $x_2$ for fixed $\nu$ as the latter are given in terms of masses and the former depends on the changing frequency. In the averaging process, one must still be careful as the formula (\ref{salakmeral})  fails for large $x$ values. The maximum possible value of $\nu$ is 0.25 which is attained if the two merging masses have exactly equal masses as stated above. In the first merger data, the merging masses are approximately close to each other ($m_1 = 29 M_\odot$, $m_2 = 36 M_\odot$), but not exactly equal, so we take $\nu =0.24$. Of course, strictly speaking, due to the mass lost in radiation, $\nu$ also changes, but it does so very slightly, so we ignore that.  As for the relevant range of the $x$ variable, even though theoretically one has $x \in [0,1)$, one can see from the definition of $x$ in (\ref{bluh}), its maximum value when computed in the Newtonian approximation, using (\ref{omega_eqn}) and (\ref{geyik}) in the first equation of (\ref{bluh}), turns out to be 1/2. Therefore, when one takes an average value of $g$, one has to consider the range $x\in [0,0.5]$. But this range is really the maximum range, and as $x$ gets closer to the upper bound, the approximation fails to be accurate as one approaches the strong gravity regime. Thus we considered the smaller interval   $x\in[0 ,0.2855]$ and the upper limit denotes the relative speed of the masses to be around a little over $v =c/2$. For this interval the average of $g$  can be found to be $\langle g \rangle = 1.156$ which suggests that one can take $\alpha_{eff}= 37/5$. {See  Fig. 1}.  For a two-parameter plot of $\alpha_{eff}$ which is relevant for other{ observed mergers that we have not discussed here, see Fig. 2.} 
\begin{figure}[h]
~~\includegraphics[scale=0.7]{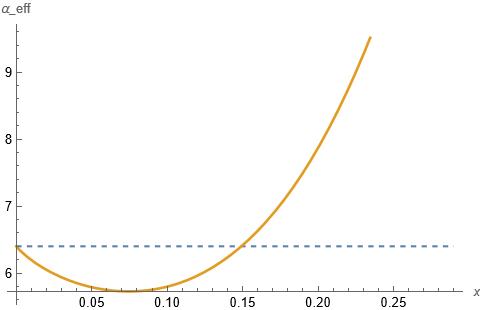}
\caption{$\alpha_{eff}$ as defined by the second equality in (\ref{salakmeral}) is plotted for  $\nu=0.24$. The horizontal line corresponds to 32/5=6.4.  The average value of $\alpha_{eff}$ for the relevant interval for the first LIGO data can be taken to be 37/5=7.4.  }
\label{fig1}
\end{figure}

\begin{figure}[h]
~~\includegraphics[scale=0.8]{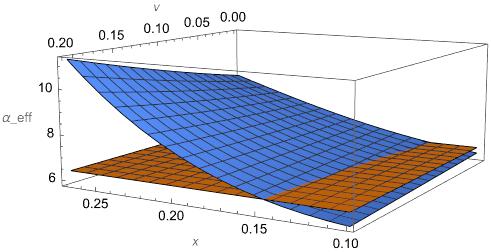}
\caption{$\alpha_{eff}$ as defined by the second equality in (\ref{salakmeral}) is plotted for  
the ranges $\nu \in [0, 1/4]$ and $x \in [0.12,0.20]$ (which is between the range stated in the text). The horizontal plane corresponds to the linearized GR's value of 32/5.  }
\label{fig2}
\end{figure}

The upshot of this discussion is that to understand the problem at a semi-Newtonian level, without going through rather cumbersome perturbative calculations, one has to choose an {\it effective} 
$\alpha_{eff} \ne \frac{32}{5}$ which should satisfy the bound (\ref{bilal1}). In the first LIGO observation \cite{LIGO_disco}, as noted in the paragraph below (\ref{bilal1}), one has $\alpha_{eff} \ge 7.0613$. We argued above that the value $\alpha_{eff} = \frac{32}{5}+1 =7.4$ is a reasonable value to choose for the first LIGO observation. Of course, staying in the noted bound (\ref{bilal1}), one can argue for some other value. We will show below that the value we take produces impressively close values to the LIGO data.

But first, let { us} recast (\ref{main_mass}) in terms of the solar mass $M_\odot$  which is more convenient:
\begin{equation}
m_{1,2}= \frac{11421 M_\odot}{f_2}\left(1 \pm \sqrt{1-\frac{2 \sqrt{2} f_2^{5/3}}{ \pi \tau \alpha_{eff}}\left(\frac{1}{f_1^{8/3}}-\frac{1}{f_2^{8/3}}\right)}\,\, \right),
\label{lz_tugce}
\end{equation}
where the frequencies are in hertz. Please note that we have changed $\alpha$ to $\alpha_{eff}$ and we shall use the value  $\alpha_{eff}=37/5$. Then using  (\ref{lz_tugce}) gives  the individual masses as $m_1 = 30 M_\odot$, $m_2 = 46 M_\odot$, $m_T = 76 M_\odot$; and the chirp mass as ${\mathcal{M}}_c = 32 M_\odot$; so we have $m_T = 2.375{\mathcal{M}}_c $ which satisfies the bound. These values are in reasonable comparison with the LIGO estimates  \cite{LIGO_disco}: $m_1 = 29 M_\odot$, $m_2 = 36 M_\odot$, $m_T = 70 M_\odot$ and ${\mathcal{M}}_c = 30 M_\odot$ and $m_T =2.3333 {\mathcal{M}}_c $.

We can also estimate the total mass carried away by the gravitational radiation. Integrating the power (\ref{pow}) over the interval $\tau$, one arrives at
\begin{equation}
 \Delta E =\int_0^\tau P(t) d t = \frac{\mu}{3}\left ( G m_T \right )^{2/3}\int_0^\tau dt  \, \omega_0^{-1/3} \,\frac {d \omega_o}{d t} = \frac{\mu}{2}\left ( \pi G m_T \right )^{2/3}(f_2^{2/3}- f_1^{2/3}),
\end{equation}
which yields a total mass of  $ \Delta E /c^2 = 3.3 M_\odot$ which is a rather remarkable agreement with the value given in  \cite{LIGO_disco}, that is $3 M_\odot$

\section{Conclusions and Discussions}

Our goal in this work was to give a simple analytical (albeit approximate) formula
 for the individual black hole masses in a merger in terms of the three measured quantities in the detector, while staying within the context of Newtonian mechanics as much as possible, and only using the radiated power formula from GR. The measured quantities are the initial ($f_1$) and maximum wave frequency ($f_2)$ and the time elapse ($\tau$) between them as measured in the detector. The formula (\ref{main_mass}) was derived in basic Newtonian physics in the spirit of \cite{Mathur,LIGO_team} with one important improvement and refinement: the power formula carried by gravitational waves cannot just be taken as the lowest order post-Newtonian term. Namely, the famous $32/5$ factor leads to complex black hole masses when Newtonian theory is employed to understand the LIGO data as was done \cite{Mathur}. The reason for this failure is easily understandable since, {due to the nonlinearity of GR}, the radiated power depends rather non-trivially on the involved masses and the frequency of the source. {The full dependence of the radiated power on the involved parameters is not known exactly. This is because, in the context of GR, there is no genuine two-body system: whenever there are two interacting, accelerating masses, there is always a third ``body'' which is the emitted gravitational radiation. As is well-known, the 3-body problem even in Newtonian physics is not exactly solvable except for certain special cases.  Thus in GR, the binary black hole system is not exactly solvable: one either works numerically or resorts to the perturbative calculations. In either case, one should define not a constant $\alpha= 32/5$, but a function as $\alpha_{eff} = \alpha_{eff}(f,m_1,m_2)$ which only yields the constant value $32/5$ at the lowest order. But a detailed discussion on the self-consistent solutions to this non-linear equation} is beyond the scope of this work which aims to discuss the problem in the semi-Newtonian paradigm. Therefore, to keep the discussion simple, one can take a constant average effective $\alpha_{eff}$ as discussed in the previous section.  For the first LIGO data, this averaging procedure gave us $\alpha_{eff} = 32/5 +1$ as suggested by the higher-order post-Newtonian calculations. The resulting estimates are in rather remarkable agreement with the observation, even though we have not considered the effect of the spin of the black holes. There is an important caveat here that must be considered when a similar approach is to be applied to other black hole merger data. First of all, in the spirit of not going through full data analysis, we estimated the initial frequency and the final frequency and the time interval between them by directly looking at the frequency versus time graphs. Of course, this type of estimation is not very accurate, if one makes a different estimate, one needs to be careful in using a reasonable $\alpha_{eff}$. This value should satisfy the bound (\ref{bilal1}).
\footnote{ For example, instead of $f_1$= 42 Hz, one can also read $f_1$= 41 Hz  from the first LIGO data. This is indeed correct, in that case, one should reconsider 
$\alpha_{eff}$.  For example,  $\alpha_{eff}$ = 38/5 satisfies the bound and yields  $m_1 = 34 M_\odot$, $m_2 = 42 M_\odot$, $m_T = 76 M_\odot$ and ${\mathcal{M}}_c = 33 M_\odot$ and $m_T =2.30 {\mathcal{M}}_c $.}

\section*{Acknowledgments} We would like to thank a very conscientious referee and a very conscientious board member of the journal whose suggestions made the manuscript more readable.

\end{document}